\documentclass[letterpaper]{article}
\usepackage{aaai}
\usepackage{times}
\usepackage{helvet}
\usepackage{courier}
\usepackage{graphicx}
\usepackage{hyperref}
\usepackage{adjustbox}
\usepackage{xcolor}
\usepackage{arabtex}
\renewenvironment{abstract}%
         {\centerline{\large\bf Abstract}%
          \begin{list}{}%
             {\setlength{\rightmargin}{0.6cm}%
              \setlength{\leftmargin}{0.6cm}}%
           \item[]\ignorespaces}%
         {\unskip\end{list}}
\usepackage{utf8}
\usepackage{ulem}

\usepackage{pgfplots}
\title{Large Arabic Twitter Dataset on  COVID-19}
\author{Sarah Alqurashi\thanks{Corresponding author: s43980127@st.uqu.edu.sa}, Ahmad Alhindi\and Eisa Alanazi\\
Center of Innovation and Development in Artificial Intelligence\\
Umm Al-Qura University\\
Makkah, Saudi Arabia\\
}
\begin{document}
\nocopyright
\maketitle

\begin{abstract}
The 2019 coronavirus disease (COVID-19), emerged
late December 2019 in China, is now rapidly spreading across
the globe. At the time of writing this paper, the number of global confirmed cases has passed two millions and half with over 180,000 fatalities. Many countries have enforced strict social distancing policies to contain the spread of the virus. This have changed the daily life of tens of millions of people, and urged people to turn their discussions online, e.g., via online social media sites like Twitter. In this work, we describe the first Arabic tweets dataset on COVID-19 that we have been collecting since January 1st, 2020. The dataset would help researchers and policy makers in studying different societal issues related to the pandemic. Many other tasks related to behavioral change, information sharing, misinformation and rumors spreading can also be analyzed.
\end{abstract}

\section*{Introduction}
On December 31, 2019, Chinese public health authorities reported several cases of a respiratory syndrome caused by an unknown disease, which subsequently became known as COVID-19 in the city of Wuhan, China.
This highly contagious disease continued to spread worldwide, leading the World Health Organization (WHO) to declare a global health emergency on January 30, 2020. 
On March 11, 2020 the disease has been identified as pandemic by WHO, and many countries around the world including Saudi Arabia, United States, United Kingdom, Italy, Canada, and Germany have continued reporting more cases of the disease \cite{world2020coronavirus}. 
As the time of writing this paper, this pandemic is affecting more than 208 countries around the globe with more than one million and half confirmed cases \cite{WHO}.

Since the outbreak of COVID-19, many governments around the world enforced different measures to contain the spread of the virus. 
The measures include travel restrictions, curfews, ban of mass gatherings, social distancing, and probably cities lock-down.
%
%
This has impacted the routine of people around the globe, and many of them have turned to social media platforms for both news and communication. 
Since the emergence of COVID-19,  Twitter platform plays a significant role in crisis communications where millions of tweets related to the virus are posted daily. 
Arabic is the official language of more than 22 countries with nearly 300 million native speakers worldwide. Furthermore, there is a large daily Arabic content in Twitter as millions of Arabic users use the social media network to communicate. For instance, Saudi Arabia alone has nearly 15 million Twitter users as of January, 2020 \cite{statista}. Hence, it is important to analyze the Arabic users' behavior and sentiment during this  pandemic. Other Twitter COVID-19 datasets have been recently proposed \cite{chen2020covid,lopez2020understanding} but with no significant content for the Arabic language. 

%

In this work, we provide the first dataset dedicated to Arabic tweets related to COVID-19. The dataset is available at \url{https://github.com/SarahAlqurashi/COVID-19-Arabic-Tweets-Dataset}. We have been collecting data in real-time from Twitter API since January 1, 2020, by tracking COVID-19 related keywords which resulted in more than 3,934,610 Arabic tweets so far. The presented dataset is believed to be helpful for both researchers and policy makers in studying the pandemic from social perspective, as well as analyzing the human behaviour and information spreading during pandemics.  

In what follows, we describe the dataset and the collection methods, present the initial data statistics, and provide information about how to  use the dataset.
\section*{Dataset Description}

We collected COVID-19 related Arabic tweets from January 1, 2020 until April 15, 2020, using Twitter streaming API and the Tweepy Python library.  We have collected more than 3,934,610 million tweets so far. In our dataset, we store the full tweet object including the id of the tweet, username, hashtags, and geolocation of the tweet. 

 We created a list of the most common Arabic keywords associated with COVID-19. 
 Using Twitter’s streaming API, we searched for any tweet containing the keyword(s) in the text of the tweet. Table \ref{tbl:keywords} shows the list of keywords used along with the starting date of tracking each keyword. Furthermore, Table \ref{tbl:hashtags} shows the list of hashtags we have been tracking along with the number of tweets collected from each hashtag. Indeed, some tweets were irrelevant, and we kept only those that were relevant to the pandemic.

\begin{table}[htb]
\begin{center}
\footnotesize
\begin{tabular}{|c|c|c|}
\hline
 \textbf{\textit{Keyword}}& \textbf{\textit{English Translation}} &\textbf{\textit{Tracing Date} } \\
\hline
\setcode{utf8}
\<الفايروس التاجي>
& Coronavirus & 2020-01-01 \\
\hline
\<كورونا> 
&Corona & 2020-01-01\\
\hline
\<كوفيد 19> 
 & COVID 19& 2020-03-01\\
\hline
\<ووهان> 
 & Wuhan& 2020-01-01\\
\hline
\<الصين> 
 & China& 2020-01-01 \\
\hline
\<وباء> 
 & Epidemic & 2020-01-22\\
\hline
\<تفشي> 
 & Outbreak & 2020-01-01\\
\hline
\<جائحة> 
 & Pandemic & 2020-01-22\\
\hline
\<كمامة >
 & Mask & 2020-01-01\\
\hline
\<كمامات> 
 & Masks & 2020-01-01\\
\hline
\<معقمات> 
 & Sterilizers& 2020-01-01 \\
\hline
\<تعقيم> 
 & Sterilization & 2020-01-01\\
\hline
\<غسل اليدين> 
 & Washing hands & 2020-01-01\\
\hline
\<العزل المنزلي>
 & Home isolation & 2020-02-01\\
\hline
\<الحجر المنزلي>
& Home quarantine & 2020-02-01\\
\hline
\<حظر التجول> 
  & Curfew & 2020-03-15\\
 \hline
\<التباعد الاجتماعي> 
  & social distancing  & 2020-04-01\\
 \hline
\< جهاز التنفس الصناعي> 
  &  ventilator & 2020-04-01\\
 \hline
\< ضيق تنفس> 
  & Shortness of breath  & 2020-04-01\\ 
\hline
\< كحة> 
  & cough  & 2020-04-01\\ 
\hline
\< حراره> 
  & temperature  & 2020-04-01\\
\hline
\< متر ونص> 
  & One and a half meters & 2020-04-01\\ 
\hline
\< فعاليات الحجر> 
  & quarantine activities  & 2020-04-01\\ 
\hline
\< الحجر الصحي> 
  & Quarantine  & 2020-04-01\\ 
\hline
\end{tabular}
\end{center}
\caption{The list of keywords that we used to collect the tweets.}
\label{tbl:keywords}
\end{table}

\begin{table}[htb]
\begin{center}
\begin{adjustbox}{width=.48\textwidth}

\begin{tabular}{|c|c|c|c|}
\hline
 \textbf{\textit{Hashtags}} & \textbf{\textit{English Translation}} &\textbf{\textit{Number of Tweets}} & \textbf{\textit{Tracing Date}} \\
\hline
\setcode{utf8}
\<الصين>
 & China & 287509 & 2020-01-01\\
  \hline
\<الصحة>
 & Health & 53186 & 2020-01-01\\
  \hline
\<كورونا>
&  Corona & 949531 & 2020-01-11 \\
\hline
\<فيروس - كورونا>
 & Coronavirus & 188214 & 2020-01-11\\
\hline
\<الكورونا>
 & Corona & 52267 &2020-01-17\\ 
  \hline
\<كورونا - الجديد>
 & New Corona&79302 & 2020-01-22\\
 \hline
\<فايروس - كورونا>
 & Corona virus &17723& 2020-01-23\\
  \hline
\<كوفيد - 19>
 & COVID 19 & 54865 & 2020-02-01\\
 \hline
\<فايروس - كورونا>
 & Corona virus &17723& 2020-02-01\\
 \hline
\<فيروس - كورونا - المستجد>
 & the emerging Corona virus & 20377 & 2020-02-22\\
\hline
\<كلنا - مسؤول>
 & We are all responsible &16768 & 2020-03-01\\
 \hline
\<ابقوا - في - منازلكم>
 & Stay at your homes &24580 & 2020-03-01\\
\hline
\<فعاليات - الحجر >
 & Quarantine activities &4396 & 2020-03-01\\
\hline
\<الحجر - المنزلي>
 & Home quarantine &44444 & 2020-03-08\\
\hline
\<حظر - تجول>
 & curfew & 39778 & 2020-03-10\\
\hline
\<خلك - في - البيت>
 & Stay at home &768196 & 2020-03-22\\
\hline
\end{tabular}
\end{adjustbox}
\end{center}
\caption{Hashtags used in collecting this dataset.}
\label{tbl:hashtags}
\end{table}

\begin{figure}[hbt]
\centering
\includegraphics[scale=.7]{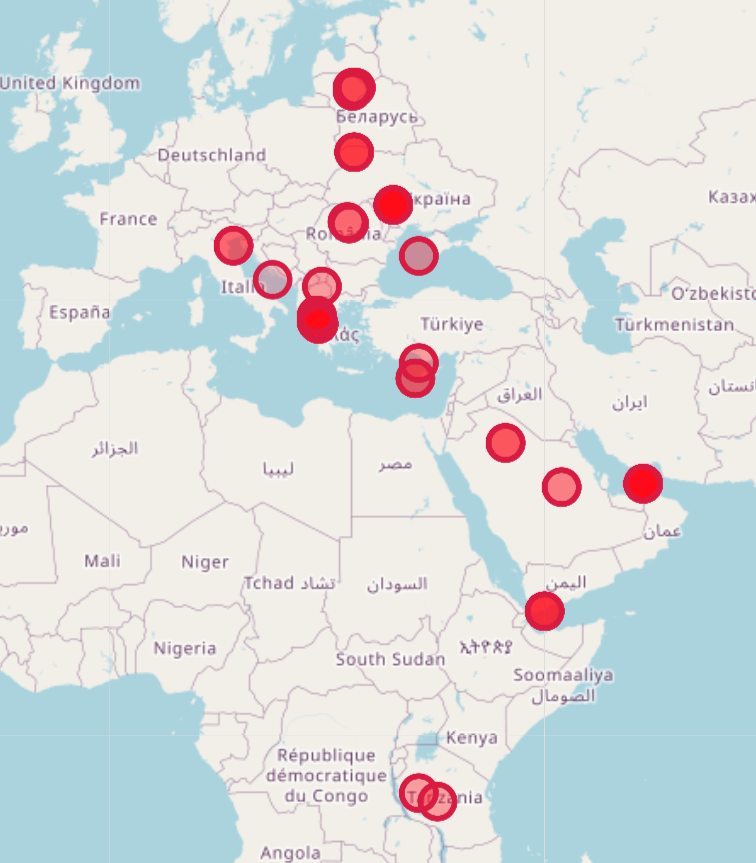}
\caption{The location of geotagged tweets}
\label{fig:geotag}
\end{figure}

A summary over the dataset is given in Table \ref{tbl:summary}. While collecting data, we have observed that the number of retweets increased significantly in late March. This is likely due to the exponential increase in confirmed COVID-19 cases worldwide, including the Arabic speaking countries.
A relatively small percentage of tweets were geotagged. Figure \ref{fig:geotag} presents the location of tweets observed as of 14 April 2020.
\begin{table}[htb]
\begin{center}
\begin{tabular}{ l c }
\hline

Number of Tweets : & 3,934,610\\
Number of Tweets with geolocation : & 219 \\
Number of original tweets:  & 3,934,235\\
Number of  retweets : & 375 \\  
The average  of daily collected tweets: & 77471 \\  
\hline
\end{tabular}
\end{center}
\caption{Summary statistics for the collected tweets}
\label{tbl:summary}
\end{table}

\section*{Dataset Access}
The dataset is accessible on GitHub at this address:
\url{https://github.com/SarahAlqurashi/COVID-19-Arabic-Tweets-Dataset}

However, to comply with Twitter’s content redistribution policy\footnote{\url{https://developer.twitter.com/en/developer-terms/agreement-and-policy}}, we are distributing only the IDs of the collected tweets. There are several tools (such as Hydrator\footnote{\url{https://github.com/DocNow/hydrator}}) that can be used to retrieve the full tweet object. We also plan to provide more details on the pre-processing phase in the GitHub page.


\section*{Future Work}
We are continuously updating the dataset to maintain more aspects of COVID-19 Arabic conversations and discussions happening on Twitter. We also plan to study how different groups respond to the pandemic and analyze information sharing behavior among the users.

\section*{Acknowledgements}
The authors wish to express their thanks to Batool Mohammed Hmawi for her help in data collection.

\bibliographystyle{aaai}
\bibliography{ref}
\end{document}